\documentstyle[12pt]{article}
\newlength{\pubnumber} 

\catcode`\@=11
\@addtoreset{equation}{section}

\def\section{\@startsection{section}{1}{\z@}{3.5ex plus 1ex minus .2ex}
 {2.3ex plus .2ex}{\large\bf}}
\def\subsection{\@startsection{subsection}{2}{\z@}{2.3ex plus .2ex}
 {2.3ex plus .2ex}{\bf}}

\renewenvironment{thebibliography}[1]
        {\begin{list}{[\arabic{enumi}]}
        {\usecounter{enumi}\setlength{\parsep}{0pt}
         \setlength{\itemsep}{0pt}
         \settowidth
        {\labelwidth}{[{#1}]}\sloppy}}{\end{list}}

\begin{document}

\setcounter{page}{1}
\rightline{\tt hep-ph/0211211}
\rightline{November 2002}
\bigskip
\begin{center}
   {\Large \bf  Beautified with Goodly Shape:\\
        Rethinking the Properties of Large Extra Dimensions\footnote{
       Invited talk given at SUSY~2002:  The 10$^{\rm th}$ International
       Conference on Supersymmetry and Unification of the Fundamental
       Interactions, held at DESY, Hamburg, Germany, June 2002.}
       \\ }
\bigskip
\bigskip
   {\large
      Keith R. Dienes \\}
\vspace{.04in}
 {\it  Department of Physics, University of Arizona, Tucson, AZ  85721  USA\\
         E-mail address:~ {\tt dienes@physics.arizona.edu}} 
\end{center}
\medskip
\begin{abstract}
  {\rm 
   Much recent attention has focused on theories with large extra
compactified dimensions.  However, while the phenomenological implications
of the volume moduli associated with such compactifications are well
understood, relatively little attention has been devoted to the shape
moduli.  In this talk, I demonstrate that non-trivial shape moduli 
can lead to  a number of effects which are
relevant not only for experimental searches
for extra dimensions, but also for the interpretation of 
experimental data if such extra dimensions are found.
This talk reports on work done partly in collaboration 
with Arash Mafi. 
   }
\end{abstract}


\def\beq{\begin{equation}}
\def\eeq{\end{equation}}
\def\beqn{\begin{eqnarray}}
\def\eeqn{\end{eqnarray}}
\def\half{{\textstyle{1\over 2}}}
\def\calM{{\cal M}}
\def\calF{{\cal F}}
\def\ie{{\it i.e.}\/}
\def\eg{{\it e.g.}\/}

\newcommand{\newc}{\newcommand}
\newc{\gsim}{\lower.7ex\hbox{$\;\stackrel{\textstyle>}{\sim}\;$}}
\newc{\lsim}{\lower.7ex\hbox{$\;\stackrel{\textstyle<}{\sim}\;$}}

\hyphenation{su-per-sym-met-ric non-su-per-sym-met-ric}
\hyphenation{space-time-super-sym-met-ric}
\hyphenation{mod-u-lar mod-u-lar--in-var-i-ant}


\def\inbar{\,\vrule height1.5ex width.4pt depth0pt}

\def\IC{\relax\hbox{$\inbar\kern-.3em{\rm C}$}}
\def\IQ{\relax\hbox{$\inbar\kern-.3em{\rm Q}$}}
\def\IR{\relax{\rm I\kern-.18em R}}
 \font\cmss=cmss10 \font\cmsss=cmss10 at 7pt
\def\IZ{\relax\ifmmode\mathchoice
 {\hbox{\cmss Z\kern-.4em Z}}{\hbox{\cmss Z\kern-.4em Z}}
 {\lower.9pt\hbox{\cmsss Z\kern-.4em Z}}
 {\lower1.2pt\hbox{\cmsss Z\kern-.4em Z}}\else{\cmss Z\kern-.4em Z}\fi}

\def\NPB#1#2#3{{\it Nucl.\ Phys.}\/ {\bf B#1} (#2) #3}
\def\PLB#1#2#3{{\it Phys.\ Lett.}\/ {\bf B#1} (#2) #3}
\def\PRD#1#2#3{{\it Phys.\ Rev.}\/ {\bf D#1} (#2) #3}
\def\PRL#1#2#3{{\it Phys.\ Rev.\ Lett.}\/ {\bf #1} (#2) #3}
\def\PRT#1#2#3{{\it Phys.\ Rep.}\/ {\bf#1} (#2) #3}
\def\CMP#1#2#3{{\it Commun.\ Math.\ Phys.}\/ {\bf#1} (#2) #3}
\def\MODA#1#2#3{{\it Mod.\ Phys.\ Lett.}\/ {\bf A#1} (#2) #3}
\def\IJMP#1#2#3{{\it Int.\ J.\ Mod.\ Phys.}\/ {\bf A#1} (#2) #3}
\def\NUVC#1#2#3{{\it Nuovo Cimento}\/ {\bf #1A} (#2) #3}
\def\etal{{\it et al.\/}}

\long\def\@caption#1[#2]#3{\par\addcontentsline{\csname
  ext@#1\endcsname}{#1}{\protect\numberline{\csname
  the#1\endcsname}{\ignorespaces #2}}\begingroup
    \small
    \@parboxrestore
    \@makecaption{\csname fnum@#1\endcsname}{\ignorespaces #3}\par
  \endgroup}
\catcode`@=12

\input epsf

\medskip
\section{Introduction:  Beautified with goodly shape}

Over the past several years, there has been an explosion of interest
in theories with large extra spacetime dimensions~\cite{Antoniadis,ADD,DDG,string}.  
Much of this
interest stems from the realization that large extra dimensions
have the potential to lower the fundamental energy scales of physics,
such as the Planck scale, the GUT scale, and the
string scale.
Indeed, the degree to which these scales
may be lowered depends on the volume of the compactified dimensions.

However, compactification geometry is not merely endowed with volume;
it is also ``beautified with goodly shape.''\/\footnote{
           W.~Shakespeare, {\it The Two Gentlemen of Verona}\/: IV, i.}
Indeed, both volume (or ``K\"ahler'') moduli and shape (or ``complex'') moduli 
are necessary in order to fully describe the geometry of compactification.  
This distinction has phenomenological
relevance because the shape moduli also play a significant role in
determining the experimental bounds on such scenarios.  
Unfortunately,
in most previous discussions of large extra dimensions, relatively little attention
has been paid to the implications of these moduli.

In this talk, I shall discuss some of the phenomenological implications of
non-trivial shape moduli~\cite{shape1,shape2,shape3}.
First, as we shall see, shape moduli can have dramatic
effects on the corresponding Kaluza-Klein (KK) spectrum~\cite{shape1}:  
they can induce level-crossings and varying mass gaps, 
they can help to eliminate light KK states,
and they can alter experimental constraints in such scenarios 
to the extent that the bounds on the largest extra dimension can be completely
eliminated.  In other words, we shall see that large extra dimensions can
be essentially invisible (even if they are flat).

Second, we shall see that shape casts ``shadows''~\cite{shape2}. 
Specifically, we shall show that non-trivial shape moduli can distort our 
experimental perceptions of compactification geometry at low energies.
Indeed, we shall find that spacetime geometry is essentially ``renormalized''
as a function of the energy scale with which it is probed.

Finally, we shall briefly discuss the effects that non-trivial shape moduli 
can have on string winding modes.  We shall show, in particular, that toroidal
compactifications exist for which all KK states as well as all winding
modes are heavier than the string scale~\cite{shape3}.  Thus, in the context of low-scale
string theories, it is possible to cross
the string scale without detecting {\it any}\/ string states   
associated with spacetime compactification.

\section{Shape matters:  A bit of background}
\setcounter{footnote}{0}

\rightline{\it To any shape of thy 
      preferment\/\footnote{W.~Shakespeare, {\it Cymbeline}\/:  I, v.}}
\bigskip

\begin{figure}[th]
\centerline{
      \epsfxsize 2.2 truein \epsfbox {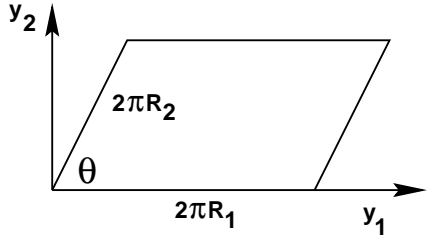}}
\caption{General two-dimensional torus with shift angle $\theta$.}
\label{torus}
\end{figure}

In order to illustrate these points, let us begin by considering the case of
compactification on a general two-torus, as illustrated in Fig.~\ref{torus}.
Such a torus is realized as the space $\IR^2$ subject to the two discrete
identifications
\beq
          \cases{  y_1 ~\to~ y_1+2\pi R_1\cr
                     y_2 ~\to~ y_2\cr}  
         ~~~~~~~~~~~\cases{  y_1 ~\to~ y_1+2\pi R_2 \cos\,\theta\cr
                     y_2 ~\to~ y_2+2\pi R_2 \sin\,\theta~.\cr}
\label{torusdef}
\eeq
Thus, the flat two-torus is specified by three parameters:  $R_1$, $R_2$, and $\theta$.
The volume of the torus is $V\equiv 4\pi^2 R_1 R_2 \sin\theta$, while
the shape parameters are $R_2/R_1$ and $\theta$.  

Most previous studies have focused on the effects of $V$ and $R_2/R_1$, 
essentially fixing $\theta=\pi/2$.  But what are the phenomenological implications
of non-trivial $\theta$?

Clearly, we need to analyze the KK wavefunctions on this space.
Demanding invariance under the torus periodicities in Eq.~(\ref{torusdef}),
we find that the KK wavefunctions take the form
\beq
     \Psi_{n_1,n_2}~\sim~
       \exp\left\lbrack i {n_1\over R_1} \left(y_1- {y_2\over \tan\theta}\right)
        ~+~ i{n_2\over R_2} {y_2\over \sin\theta} \right\rbrack~,~~~~~~~ n_i\in\IZ~.
\label{wavefunctions}
\eeq
Applying the (mass)$^2$ operator $- (\partial^2 /\partial y_1^2 +
            \partial^2 /\partial y_2^2)$,
we then find the KK masses
\beq
   \calM_{n_1,n_2}^2 ~=~ {1\over \sin^2\theta} \left(
         {n_1^2\over R_1^2} + {n_2^2\over R_2^2} -
         2 {n_1n_2\over R_1R_2} \cos\theta\right)~.
\label{KKmasses}
\eeq
Note that these masses are no longer invariant under $n_1\to -n_1$ or $n_2\to -n_2$
individually unless combined with $\theta\to \pi-\theta$.  We can therefore restrict
our attention to values of $\theta$ in the range $0<\theta\leq \pi/2$ without loss of generality.

\begin{figure}
\centerline{
      \epsfxsize 3.2 truein \epsfbox {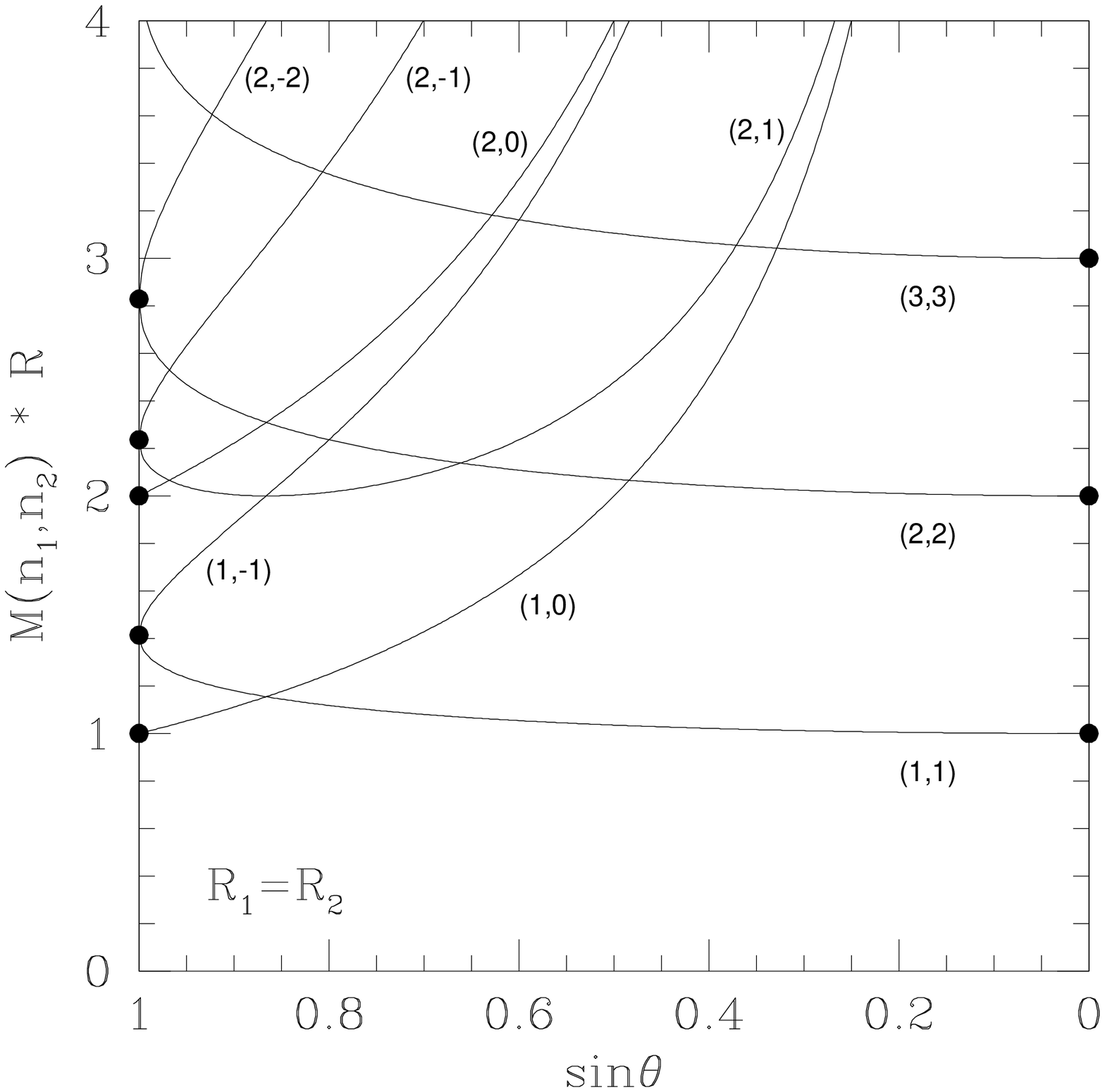}
      \epsfxsize 3.2 truein \epsfbox {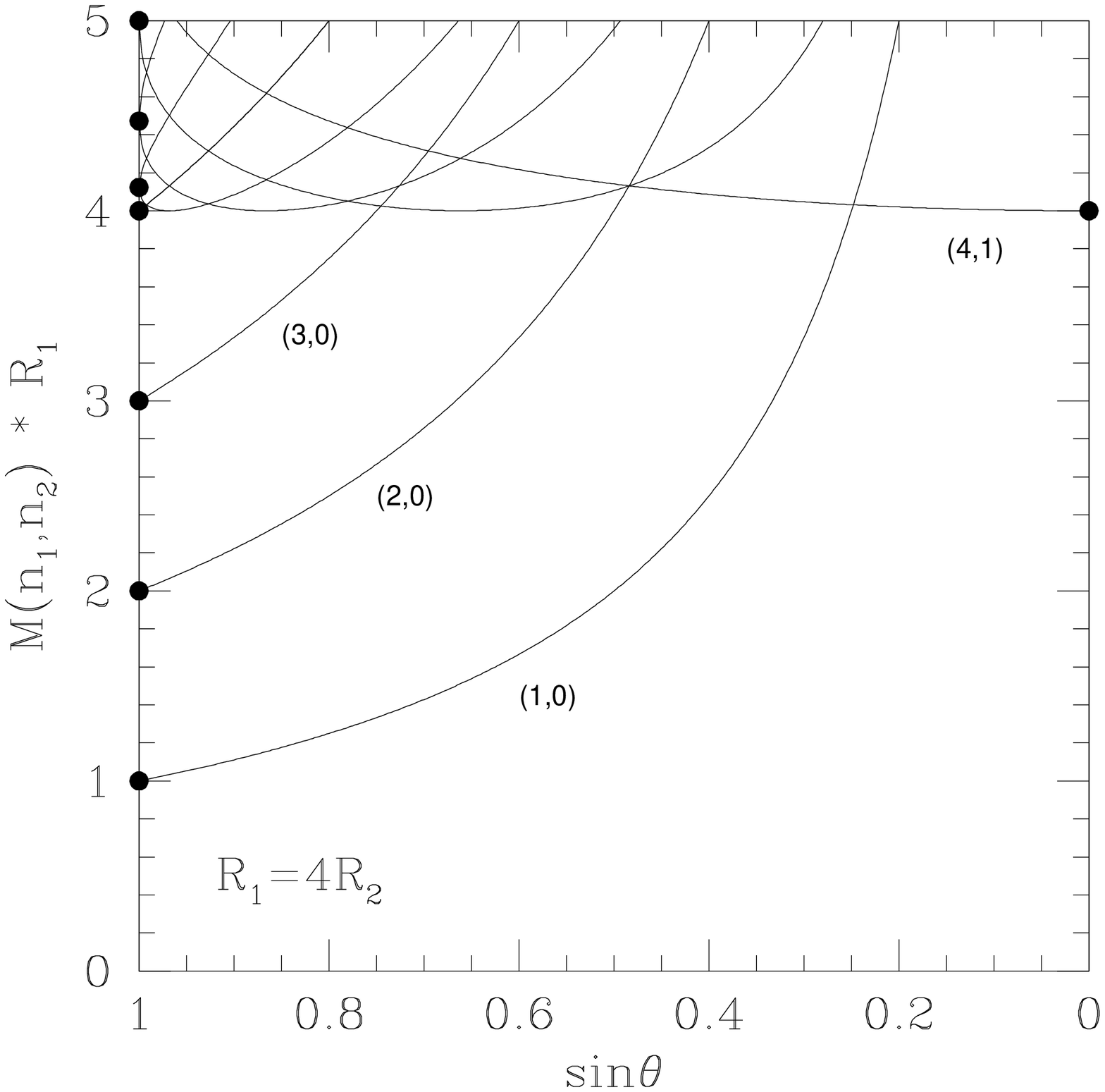} }
\centerline{
      \epsfxsize 3.2 truein \epsfbox {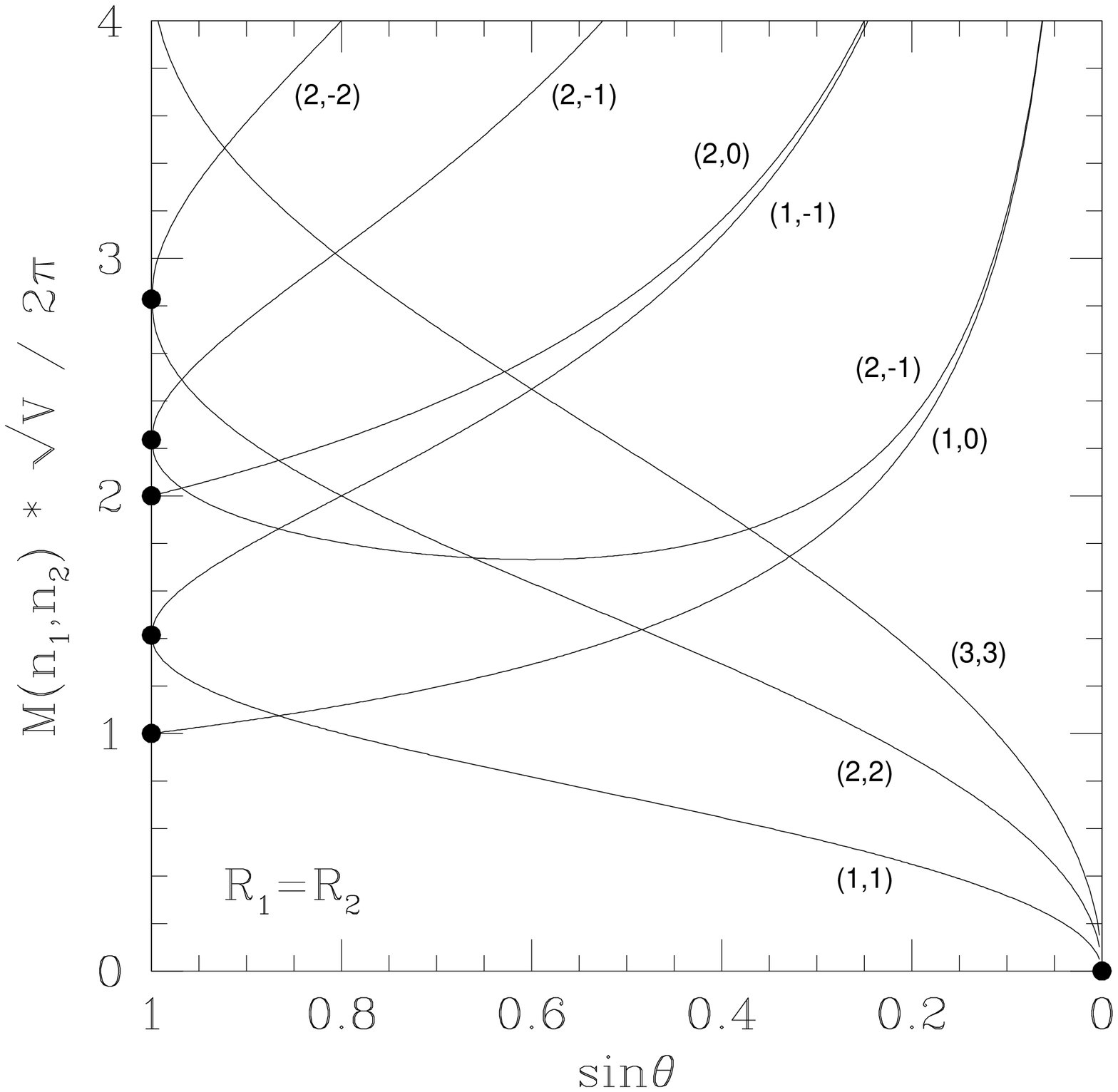}
      \epsfxsize 3.2 truein \epsfbox {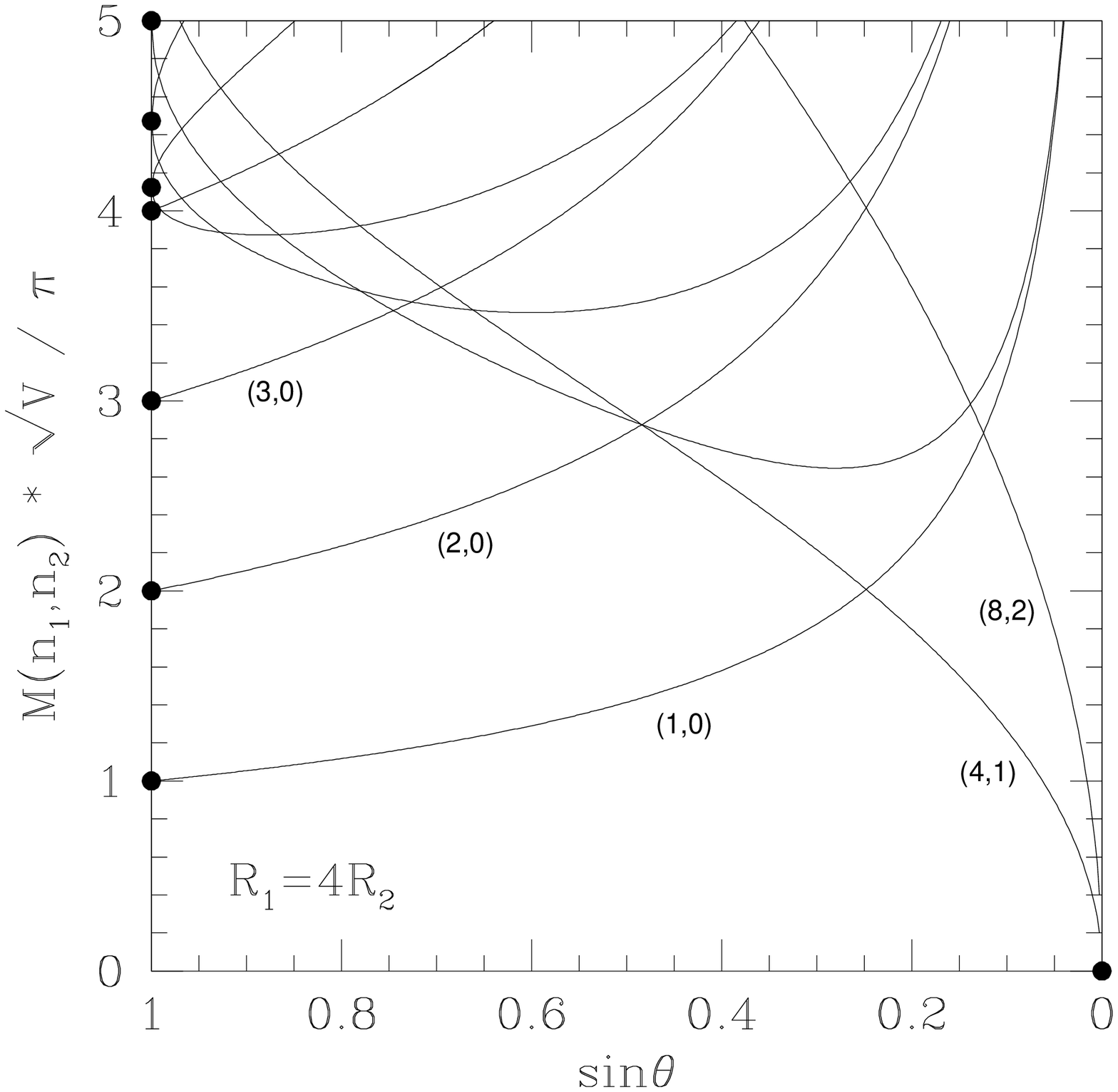} }
\caption{The masses of the lowest-lying KK states $(n_1,n_2)$
        as functions of the shape parameter $\theta$
        for  $R_1=R_2\equiv R$ (upper left plot) and
             $R_1=4 R_2$ (upper right plot), holding 
        $R_1$ and $R_2$ fixed.  The lower plots 
       represent the same
       scenarios where we vary $\theta$ holding
       both the volume of the torus and the ratio $R_2/R_1$ fixed.}
\label{plots}
\end{figure}

The case with $\theta=\pi/2$ corresponds to the usual ``rectangular'' torus,
but we are interested here in the behavior of the KK spectrum
as $\theta$ is varied while
holding the radii ($R_1,R_2$) fixed. 
We then find dramatic changes in the KK spectrum,
as illustrated in Fig.~\ref{plots} (upper plots) for the cases
with $R_1=R_2$ and $R_1=4 R_2$.
We observe that in general, 
the {\it mass gap}\/ $\mu$ (defined
as the mass of the first excited KK state)
depends on $\theta$, and tends to {\it increase}\/ 
as $\theta\to 0$.
In the case of KK gravitons, 
this increase in the mass gap implies that 
the deviations from Newtonian gravity will be {\it exponentially
suppressed}\/ as $\theta\to 0$, even though the radii of the
extra dimensions are held fixed. 

Another possibility, of course, is to hold the {\it volume}\/ of the torus fixed 
(as well as the ratio $R_2/R_1$) while we vary $\theta$.
We then find the results illustrated in Fig.~\ref{plots} (lower plots).
In this case, the mass gap can vary, but need not 
necessarily grow as $\theta\to 0$.

\section{The collapsing torus:  A closer look}
\setcounter{footnote}{0}

\rightline{\it  Changed to a worser shape thou canst not 
      be.\/\footnote{W.~Shakespeare, {\it King Henry VI, Part I}\/:  V, iii.}}

\bigskip

The limit as $\theta\to 0$ is interesting and requires a closer look.
It turns out~\cite{shape1} that the resulting behavior of the lightest KK modes 
depends critically on whether the ratio $R_2/R_1$ is rational or irrational.

Let us first consider the case when the radii are held fixed.
If $R_2/R_1$ is rational, so that we can write $R_2/R_1=p/q$ where $p,q\in\IZ$
in lowest form, then in the $\theta\to 0$
limit we find that the two torus periodicities in Eq.~(\ref{torusdef}) collapse
to form a single periodicity corresponding a circle of radius $R\equiv R_1/q = R_2/p$.
Indeed, the states with $(n_1,n_2)= k(p,q)$ where $k\in \IZ$ remain finitely
massive in this limit, asymptotically becoming the circle KK states with  
masses $k/R$, while all other KK states become infinitely heavy and decouple
from the spectrum.
This behavior is shown in Fig.~\ref{plots} (upper plots).
If $R_2/R_1$ is {\it irrational}\/, however, 
then $p$ and $q$ are essentially infinite, resulting in a circle with
vanishing radius;  {\it all}\/ KK states become infinitely heavy as $\theta\to 0$.
This reflects the incompatibility of the two toroidal periodicities
which grows increasingly severe as the torus collapses.

When the volume is held fixed, by contrast, the radii must grow to compensate as $\theta\to 0$.
These growing radii therefore provide an extra tendency towards making the KK states
lighter than they would have been if the radii (and not the volume) had been held fixed.
Indeed, if $R_2/R_1$ is rational, we see from Fig.~\ref{plots} (lower plots) 
that the circle-compactified states now become {\it massless}\/ as $\theta\to 0$.

A more interesting situation occurs when $R_2/R_1$ is irrational.
In this case, the incompatibility of the toroidal periodicities which forces
the KK states to become infinitely heavy is balanced against the growing
radii which tend to push the KK states to become massless.
The outcome of this competition turns out to 
depend~\cite{shape1} on whether 
$R_2/R_1$ is transcendental (such as $\pi$ or $e$) or merely algebraic [such
as $\sqrt{2}$ or the Golden Mean $\gamma\equiv (1+\sqrt{5})/2$].
If $R_2/R_1$ is transcendental, the tendency towards masslessness wins,
and we again obtain massless KK states as $\theta\to 0$.

However, if $R_2/R_1$ is algebraic, 
it turns out that there is a {\it lower bound}\/ on 
the masses of the KK states as $\theta\to 0$ with the 
compactification volume $V$ held fixed~\cite{shape1}.
As $\theta\to 0$, one finds that
              $V \calM^2 \geq 4\pi^2 A$
where the lower bound $A$
depends on certain number-theoretic
properties of the ratio $R_2/R_1$, but does not depend directly on its actual value.
Specifically, there exists an infinite set of algebraic irrational
values of $R_2/R_1$ which all lead to the same value of $A$;
moreover, for any value of $A$, the corresponding values of $R_2/R_1$
include values with arbitrarily large magnitudes.
Thus, in this sense, we see
that the lower bound on the 
masses of the excited KK states is essentially {\it independent}\/ of 
the value of the ratio $R_2/R_1$ in the $\theta\to 0$ limit, even though
both the compactification volume $V$ and the ratio $R_2/R_1$ are held fixed.

\section{Making large, flat extra dimensions invisible}
\setcounter{footnote}{0}

\rightline{\it Thy shape invisible retain thou
      still.\/\footnote{W.~Shakespeare, {\it The Tempest}\/:  IV, i.}}
\bigskip

This last observation has an important consequence.
Let us assume that $R_2\geq R_1$ without loss of generality,
and consider the experimental bounds that ordinarily 
restrict the possible sizes of extra dimensions.

In the case of a rectangular torus (for which $\theta=\pi/2$),
it is straightforward to see that 
$V \calM^2 \geq 4\pi^2 R_1/R_2$. 
However, the fact that we have not yet detected extra dimensions
experimentally 
provides an experimental bound on the 
lightest possible KK state:  $\calM \geq M_{\rm expt}$
where $M_{\rm expt}$ is the experimental bound.
Combining these two results, and recognizing that $V\equiv 4\pi^2 R_1R_2$,
we obtain the constraint
$R_2 \leq M_{\rm expt}^{-1}$.
In other words, for rectangular tori, {\it the experimental bounds on
the non-observation of KK states become a bound on the size 
of the largest extra dimension}\/, independent of the compactification
volume.

This situation changes dramatically in the $\theta\to 0$ limit. 
As we have seen, in this limit the KK mass gap satisfies
$V \calM^2 \geq 4\pi^2 A$
where $A$ is independent of the specific value of $R_2/R_1$.
Again imposing the experimental constraint
$\calM \geq M_{\rm expt}$,
we see that we now obtain a bound directly on the compactification
volume:  $V \leq 4\pi^2 A (M_{\rm expt})^{-2}$.
In other words, the experimental bound becomes independent of the
size of the largest extra dimension!
Thus, a large, flat extra dimension can be essentially {\it invisible}\/
in the $\theta\to 0$ limit, even though the volume of compactification
is held fixed.
This result is discussed further in Ref.~\cite{shape1}.

\section{Correspondence relations:\\
Is one torus is merely the base of another?}
\setcounter{footnote}{0}

\rightline{\it Wherefore base?}
\rightline{\it When my dimensions are as well compact,}
\rightline{\it My mind as generous, and my shape as 
      true.\/\footnote{W.~Shakespeare, {\it King Lear}\/:  I, ii.}}
\bigskip

\begin{figure}[t]
\centerline{\epsfxsize 2.8 truein \epsfbox {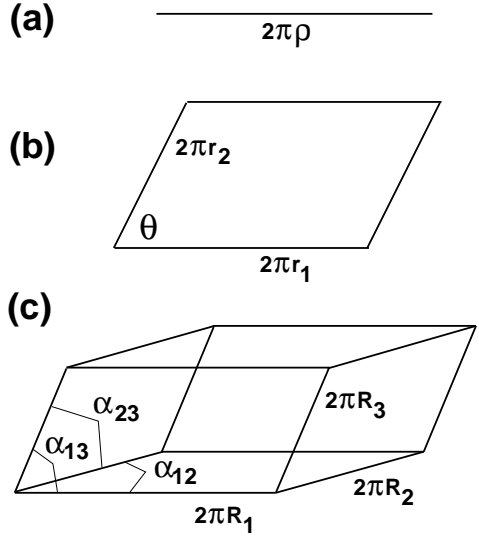}}
\caption{General one-, two-, and three-dimensional tori with arbitrary
     shape angles.}
\label{tori}
\end{figure}

Let us now consider general tori 
in one, two, and three dimensions.
These tori are illustrated in Fig.~\ref{tori}.
The one-dimensional torus, of course, is nothing but a circle
and has no corresponding shape moduli.
By contrast, the two- and three-dimensional tori are described
not only by radii but also by the shift angles $\theta$
and $\alpha_{ij}$ which mix the periodicities associated 
with translations along the corresponding directions.
Note that we shall now use the symbol $\rho$ to indicate
the radius of the one-torus;  lowercase $r_1,r_2$ to indicate the
radii of the two-torus, with $\theta$ serving as the shape angle;
and uppercase $R_1,R_2,R_3$ to indicate the radii of the three-torus,
with $\alpha_{12},\alpha_{13},\alpha_{23}$ denoting the corresponding shape angles.

Our goal is to study the extent to which various low-energy observers
can determine the shapes of these tori by studying their associated
KK spectra.
Towards this end, let us assume that
the ``true'' compactification geometry is given by the three-torus
shown in Fig.~\ref{tori}(c).  Furthermore, let us assume that
there is a hierarchy of length scales such that $R_3\ll R_2\ll R_1$.
For example, $R_3$ might be near the Planck scale, while $R_1$ might
be at the inverse TeV scale (or even the millimeter scale)
and $R_2$ might be at some intermediate
scale.
Of course, to a high-energy observer with access to energies
$E_{\rm max}\gg {\cal O}(R_3^{-1})$, the KK spectrum
will reveal the presence of all three dimensions of the torus.
Such an observer can then determine all three radii $R_i$ and
shape angles $\alpha_{ij}$ through a detailed spectral analysis
of the KK states.
However, for an observer with
access to only intermediate energies
${\cal O}(R_2^{-1}) \ll E_{\rm max} \ll {\cal O}(R_3^{-1})$,
the third dimension will be inaccessible;  the compactification
manifold would then appear to be a two-torus, as illustrated
in Fig.~\ref{tori}(b).
Finally, for the low-energy observer with access to energies
${\cal O}(R_1^{-1}) \ll E_{\rm max} \ll {\cal O}(R_2^{-1})$,
only one dimension worth of KK states will be accessible.
Such an observer would then conclude that the compactification
manifold is merely a circle, as illustrated in Fig.~\ref{tori}(a).

This change in the effective dimensionality of the compactification space
is obvious, and is not our focus in this talk.  However, given the
hierarchy $R_3\ll R_2\ll R_1$,
it is natural to expect that the intermediate-energy observer
would experience a two-torus
whose parameters $(r_1,r_2,\theta)$ are related to
the underlying parameters $(R_i,\alpha_{ij})$ of the
three-torus via the relations
\beq
          {r_1 = R_1} ~,~~~~~
                  r_2 = R_2 ~,~~~~~
                 \theta = \alpha_{12}  ~.
\label{wrong1}
\eeq
After all, at energies much below $R_3^{-1}$, we expect all remnants
of the third dimension to vanish, so that the two-torus experienced
by the intermediate-energy observer is merely the {\it base}\/ of the original
three-torus in Fig.~\ref{tori}(c).
Likewise, it is natural to expect that the lowest-energy observer
would perceive a circle with radius $\rho=r_1$, which by Eq.~(\ref{wrong1})
implies
\beq
      \rho ~=~  r_1 ~=~ R_1  ~.
\label{wrong2}
\eeq
Once again, this would be the na\"\i ve
expectation, given that we have only sufficient energy
to probe the largest dimension of
the original three-torus.

\section{Shadowing}
\setcounter{footnote}{0}

\rightline{\it To worship shadows and adore false 
      shapes\/\footnote{W.~Shakespeare, 
          {\it The Two Gentlemen of Verona}\/:  IV, ii.}}
\bigskip

We shall now demonstrate that the above relations 
in Eqs.~(\ref{wrong1}) and (\ref{wrong2})
are incorrect, even in the presence of a large hierarchy $R_3\ll R_2\ll R_1$,
and must be replaced by relations which are far more non-trivial.
Indeed, as we shall see,
the relations in Eqs.~(\ref{wrong1}) and (\ref{wrong2})
hold only when the shape moduli are ignored
(\ie, when all shape angles are taken to be $\pi/2$).
In the presence of non-rectangular shape angles, by contrast,
we shall see that these relations completely fail to describe the
process by which small extra dimensions can be ``integrated
out'' when passing to larger and larger length scales.
We stress that this failure occurs {\it no matter how small}\/ the
smallest radii become.  As we shall see, this
failure arises because of a phenomenon called ``shadowing''~\cite{shape2} 
which can have dramatic phenomenological consequences at low energies.

It is straightforward to determine the correct correspondence relations
by comparing the KK spectra in each case.
In the case of the circle in Fig.~\ref{tori}(a), the KK
spectrum is given by $\calM^2 = n_1^2 /\rho^2$ where $n_1\in\IZ$.
By contrast,
for the general two-torus shown in Fig.~\ref{tori}(b), 
we have already seen in Eq.~(\ref{KKmasses})
that the KK spectrum is given by
\beq
        \calM^2 ~=~  {1\over \sin^2\theta}
         \left\lbrack
                  \sum_{i=1}^2 {n_i^2 \over r_i^2}
             -   \sum_{i\not= j}  {n_i n_j \over r_i r_j} \cos\theta
          \right\rbrack~
\label{KKtwo}
\eeq
where $n_i\in \IZ$.
Finally, for the general three-torus shown in Fig.~\ref{tori}(c), the
KK spectrum is given by
\beq
        \calM^2 ~=~ {1\over K}\,
          \left\lbrack
           \sum_{i=1}^3 {n_i^2 \over R_i^2} s_{jk}^2
           - \sum_{i\not= j}   {n_i n_j \over R_i R_j}
             \left(c_{ij} - c_{ik} c_{jk} \right)
                \right\rbrack ~
\label{KKthree}
\eeq
where $k\not= i,j$, where
$c_{ij}\equiv \cos\alpha_{ij}$
and
$s_{ij}\equiv \sin\alpha_{ij}$,
and where $K$ [the dimensionless squared volume of the parallelepiped
in Fig.~\ref{tori}(c)]
is given by
\beq
         K \equiv 1-  \sum_{i<j} c_{ij}^2 + 2 \prod_{i< j} c_{ij}
           =
            \sum_{i<j} s_{ij}^2 + 2 \left(\prod_{i< j} c_{ij} -1 \right)~.
\eeq

Given these results, we can now determine the appropriate correspondence relations.
In deriving these relations, we shall assume a hierarchy $R_3\ll R_2\ll R_1$
so that we can successively integrate out small extra dimensions
when passing to larger length scales.
Our procedure will be to disregard all KK
states whose masses exceed the appropriate reference energy
(either high, intermediate, or low) and are therefore inaccessible to
the corresponding observer.

The observer at highest energy clearly sees three dimensions worth
of KK states, and deduces the ``true'' geometry of the
compactified space by comparing the measured KK masses
with Eq.~(\ref{KKthree}).
However, the observer at intermediate energy cannot perceive excitations
in the $R_3$ direction, since functionally $R_3\to 0$ for this
observer.  His attention is therefore restricted
to states with $n_3=0$, and he attempts a spectral analysis
of the remaining states.
This leads to the identifications
\beqn
       {1\over \sin^2\theta} {1\over r_i^2} &=& {s_{jk}^2\over K R_i^2}~~~~~ (i=1,2)\nonumber\\
       {\cos\theta\over \sin^2\theta} {1\over r_1 r_2} &=&
           { c_{12} - c_{13}c_{23} \over K R_1 R_2}~.
\eeqn
This observer therefore deduces that the compactified space
is a two-torus parametrized by $(r_1,r_2,\theta)$
given by
\beqn
   \cases{ \displaystyle{r_i}  &=  $\displaystyle{s_{i3}  R_i}$   \cr
           \displaystyle{\cos\theta}  & =
             $\displaystyle{ (c_{12} - c_{13}c_{23}) / s_{13} s_{23} }$~.\cr}
\label{right1}
\eeqn
Note that both the radii $r_i$ and the shape angle $\theta$ are affected, leading
to apparent values for $(r_1,r_2,\theta)$ which are not present in
the original three-torus.

The lowest-energy observer, by contrast, misses the $n_2$ excitations
as well.
Upon comparing with the circle KK spectrum $\calM^2 = n_1^2/\rho^2$,
he therefore concludes that the compactified space is a circle
of radius
\beq
        \rho ~=~ (\sin\theta) r_1 ~=~ {\sqrt{K} \over  s_{23}} \,R_1~.
\label{right2}
\eeq
Once again, this radius does not correspond to any periodicity
in the original three-torus.

Mathematically, these results
reflect the geometric ``shadows''
that successive smaller extra dimensions cast onto
the larger extra dimensions when they are integrated out.
As such, they indicate that
the low-energy observer can see only those ``projections''
of the compactification space which are perpendicular to
the extinguished dimensions.
But given the assumed large hierarchy of length scales,
the physical implications of this shadowing effect are
rather striking.
A small extra spacetime dimension --- even one no larger
than the Planck length! --- is able to cast a huge shadow
over all other length scales and their associated dimensions,
completely distorting our low-energy perception
and interpretation of the compactification geometry.
Indeed, the physics which we would normally associate with the
Planck scale (such as the angles that parametrize the shape
of the Planck-sized extra dimensions relative to the
larger extra dimensions) fail to decouple
at low energies.

Of course, no observer at any energy scale can use this shadowing
phenomenon in order to deduce the existence of an extra spacetime
dimension beyond his own energy scale.
Nevertheless, the observer's {\it interpretation}\/ of
that portion of the compactification geometry accessible to him
is completely distorted, leading him to
deduce geometric radii and shape angles that have no basis
in reality.
Since the existence of an even
smaller extra dimension beyond those already perceived
can never be ruled out,
this shadowing effect implies that {\it one can
never know the ``true'' compactification geometry}\/.
Even when light KK states are detected and successful
fits to the KK mass formulae are obtained via
spectral analyses,
the presence of further additional dimensions with appropriate shape
moduli can always reveal the previous successes to have been
illusory.

We are not claiming that no ``true'' compactification
geometry can ever exist.  Indeed, if one takes the predictions of string
theory seriously, then there is ultimately a true, maximum number
of compactified dimensions, with associated radii and
shape moduli.  However, as an {\it experimental}\/ question,
one can never be satisfied concerning the true number of extra
dimensions.  Thus, our result implies that one can correspondingly never
be certain of the nature of whatever compactification geometry is ultimately
discovered.  In this sense, the concept of a ``true'' compactification
geometry does not exist.

\section{Examples of shadowing}
\setcounter{footnote}{0}

\rightline{\it Examples gross as earth exhort
      me\/\footnote{W.~Shakespeare, {\it Hamlet}\/:  IV, iv.}}
\bigskip

We begin with a simple example:
suppose $R_1^{-1}= 1 ~{\rm TeV}$,  
        $R_2^{-1}= 10^{11} ~{\rm GeV}$,
        $R_3^{-1}= 10^{19} ~{\rm GeV}$,
with $\alpha_{12}  =90^\circ$ and
               $\alpha_{13} = \alpha_{23} =60^\circ$. 
How is this torus perceived at lower energies?
Using the above results, 
we see that at intermediate energies this appears to be a two-torus with
 $r_1^{-1}\approx 1.155 ~{\rm TeV}$,
 $r_2^{-1}\approx 1.155 \times 10^{11} ~{\rm GeV}$,
and $\theta\approx 71^\circ$.
Finally,  at lowest energies we observe a single one-dimensional
circle with radius 
$\rho^{-1} \approx 1.225~{\rm TeV}$.
Note that the distortions are not  large in this example.
However, this distortion for the size of the largest extra dimension
persists over {\it eight orders of magnitude}\/
in this example, and it does disappear even as the smallest
dimension(s) are taken to zero size!

Let us now consider more dramatic examples of this ``shadowing'' phenomenon.
Suppose the original three-torus 
has $R_1=R_2$ and $R_3^{-1}= 10^{19}$~GeV,
with $\alpha_{12}=90^\circ$, $\alpha_{13}=60^\circ + \epsilon$,
and $\alpha_{23}=30^\circ + \epsilon$, where $\epsilon\ll 1$.
In this configuration, the third periodicity associated with the Planck-sized extra dimension
is nearly in the plane formed by the periodicities of the two larger extra dimensions.
After integrating out the Planck-sized extra dimension,
we apparently observe a two-torus with $r_1/r_2=\sqrt{3}$ and $\theta\approx \epsilon$.
Thus the ``squashing'' of the original Planck-sized
         extra dimension relative to the large dimensions
         is perceived by a low-energy observer
         as a squashing of the two large
         dimensions with respect to each other!
Note, moreover, that this also reproduces the preconditions
(algebraic irrational ratio of radii with collapsing angle)
that were required for the ``invisibility'' mechanism discussed earlier in Sect.~4.

Now let us suppose that the original three-torus has
$R_1=R_2$ and $R_3^{-1}= 10^{19}$~GeV,
with $\alpha_{12}=90^\circ$ and $\alpha_{13}\ll \alpha_{23}$.
We then find that the effective two-torus at lower energies apparently
has $r_1\ll r_2$.  Thus,
an apparent physical hierarchy in ``large'' radii has been
              {\it generated}\/
              by the effects of a Planck-scale extra dimension!

Finally, let us consider a case with
$R_1=R_2$ and $R_3^{-1}= 10^{19}$~GeV,
where $\alpha_{12}=60^\circ$ and $\alpha_{13}=\alpha_{23}= 45^\circ$.
We then find that the resulting two-torus at low energies 
appears to be rectangular:  $\theta=90^\circ$.
Thus,  even {\it trivial}\/ shape moduli can be low-energy
illusions produced by shadowing.

\section{The ``renormalization'' of compactification geometry}
\setcounter{footnote}{0}

\rightline{\it  Yes, my lord, you see them 
 perspectively\/\footnote{W.~Shakespeare, {\it King Henry V}\/:  V, ii.}}
\bigskip

What are we to make of these results?
Clearly, as we go to higher and higher energies and discover
    additional spacetime dimensions, our description of the compactification
          manifold changes.
Our perception of quantities
       such as the radii and shape moduli associated with the largest
    (and experimentally accessible) extra dimensions continually evolves
       as a function of the energy
        with which we probe this manifold --- even though the largest
        extra dimensions are already detected and their geometric properties
        are already presumed known.
Thus, the apparent compactification geometry is not fixed at all, but
         rather undergoes {\it renormalization}
         much like other ``constants'' of nature.

This may seem to be a rather unusual way to interpret these results.
However, compactifications on these different tori constitute  
a series of different effective theories.
As we cross the energy thresholds associated with the
different extra dimensions, our effective theory changes.
In this sense, then, our correspondence relations are ``matching conditions'' or
             ``threshold effects'' which describe the same physics at
              different energy scales.
Indeed, such relations merely reflect the requirement that the physical
              KK spectrum remain invariant under change in our effective
              field theories with different cutoffs.
However, this is nothing but renormalization.
Indeed, we can imagine extrapolating our calculations
         to incorporate a continuing series of hierarchies
         corresponding to a continuing series of extra dimensions.
The corresponding series of matching conditions
            would then constitute a renormalization group ``flow.''

We stress that this 
is a purely classical effect.  It is triggered
entirely by non-trivial shape moduli, and thus is a property of
geometry itself.
However, as an {\it experimental} question, there can always be smaller
which have not yet been discovered.
We see, then, that there is no such thing as ``true'' 
compactification geometry.
Indeed, even if we think we have discovered an extra dimension with a given
size, the subsequent discovery of an additional extra dimension 
which is even smaller requires that we re-evaluate the size of the extra
dimension that had already been discovered and measured.
Further discussion of these points can be found in Ref.~\cite{shape2}.

\section{KK states, winding states, and the string scale} 
\setcounter{footnote}{0}

\rightline{\it The strings, my lord, are
      false.\/\footnote{W.~Shakespeare, {\it Julius Caesar}\/:  IV, iii.}}
\bigskip

Finally, if string theory is the ultimate theory, then spacetime
compactification
should produce not only KK states, but also winding states.
Ordinarily, the masses of KK states and winding states play a
reciprocal role:  if the lightest KK states are lighter than the string
scale, then the corresponding winding states are necessarily
heavier than the string scale.  Similarly, the reverse situation in which
the lightest KK states are heavier than the string scale
ordinarily results in winding states which are lighter than the
string scale (and is equivalent to the previous configuration
as a result of $T$-duality).
The expectation, then, is that {\it either}\/ the KK
states {\it or}\/ the winding states must
have masses at or below the string scale.
Thus, it would seem to be
impossible to cross the string scale without seeing at
least {\it some}\/ states (either KK or winding) associated
with the spacetime compactification.

However, this expectation also fails when non-trivial shape moduli
are involved~\cite{shape3}.
Specifically, it can be shown~\cite{shape3}
that it is possible for the string scale to be
simultaneously {\it lighter}\/ than {\it all}\/ the KK states
as well as {\it all}\/ the winding states.
Thus, in such theories, it is possible to cross the string scale
without seeing a single resonance associated with the spacetime
compactification!
Needless to say, this can therefore give rise to a
low-energy phenomenology which
is markedly different from that of the usual
KK effective field theories.
Further details and discussion can be found in Ref.~\cite{shape3}.

\section{Conclusions and open questions} 

\rightline{\it  Beauteous as ink:  a good
      conclusion.\/\footnote{W.~Shakespeare, {\it Love's Labour's Lost}\/:  V, ii.}}
\bigskip

We have seen that non-trivial shape moduli have the potential to
produce some surprising and counter-intuitive results.
In Sect.~4, for example, we have shown that non-trivial
shape moduli have the potential to make certain large, flat extra dimensions
essentially invisible.
This feature
can therefore be exploited to alleviate (or perhaps even eliminate)
many of the experimental bounds
(such as those from table-top
Cavendish experiments, colliders, and astrophysical phenomena)
which constrain such higher-dimensional theories.
Likewise, we have seen that non-trivial shape moduli 
lead to a shadowing phenomenon whereby small, Planck-sized extra
dimensions can profoundly alter our perception of larger, TeV-sized
(or even millimeter-sized) extra dimensions. 
These results provide important lessons for interpreting the
results of potential KK discoveries.

Needless to say, these results concerning shape moduli leave many questions
unanswered.  What sets the values of these shape moduli~\cite{shape1}?  
What are the effects of non-trivial shape moduli in more general toroidal compactifications
with  background torsion fields~\cite{shape3}, or in more 
complicated non-toroidal geometries?
How do modular symmetries affect these results~\cite{shape2}, and 
how might these results be interpreted within the context of string theory
when non-trivial vacuum energies and GSO projections are present~\cite{shape3}?
How can the
degrees of freedom associated with these non-trivial shape moduli be exploited
in order to produce phenomenologically interesting theories,
and how do they affect current collider (and Cavendish) bounds on extra dimensions?
How might theories with non-trivial shape moduli be deconstructed~\cite{deconst}?
If shape moduli can have such dramatic effects on KK masses,
what are their effects on couplings and scattering amplitudes~\cite{toappear}?
How do quantum effects contribute to the renormalization of compactification
geometry?
And if compactification geometry is renormalized, to what extent
is spacetime geometry knowable at all?  
Indeed, what are the analogues of the renormalization-group invariants?
Finally, can we use the phenomenon of Planck-scale ``shadowing'' 
in order to develop a new approach
for attacking the cosmological constant problem?  
Clearly, these and other questions await further study.

\section*{Acknowledgments}
\setcounter{footnote}{0}

\rightline{\it Go, presently inquire, and so will I,}
\rightline{\it Where money is, and I no question 
      make.\/\footnote{
    W.~Shakespeare, {\it  Merchant of Venice}\/:  I, i.}}
\bigskip

This work is supported in part by the National Science Foundation
under Grant~PHY-0071054,
and by a Research Innovation Award from Research Corporation.
I wish to thank the SUSY 2002 organizers for arranging such
an interesting conference.
I also wish to thank Arash Mafi for an enjoyable collaboration
on some of the work described in this talk.

\bigskip
\bibliographystyle{unsrt}

\bigskip

\begin{thebibliography}{99}


\bibitem{Antoniadis}
       I.~Antoniadis, Phys.\ Lett.\ B {\bf 246} (1990) 377;
        I.~Antoniadis, K.~Benakli and M.~Quiros,
        Phys.\ Lett.\ B {\bf 331} (1994) 313
         [arXiv:hep-ph/9403290].

\bibitem{ADD}   N.~Arkani-Hamed, S.~Dimopoulos and G.~Dvali,
           Phys.\ Lett.\ B {\bf 429} (1998) 263
          [arXiv:hep-ph/9803315];
            I.~Antoniadis {\it et al.},
             Phys.\ Lett.\ B {\bf 436} (1998) 257
              [arXiv:hep-ph/9804398].

\bibitem{DDG}
         K.R.~Dienes, E.~Dudas and T.~Gherghetta,
           Phys.\ Lett.\ B {\bf 436} (1998) 55
          [arXiv:hep-ph/9803466];
           Nucl.\ Phys.\ B {\bf 537} (1999) 47
               [arXiv:hep-ph/9806292];
           arXiv:hep-ph/9807522.

\bibitem{string}
          E.~Witten, Nucl.\ Phys.\ B {\bf 471} (1996) 135
          [arXiv:hep-th/9602070];
          J.D.~Lykken, Phys.\ Rev.\ D {\bf 54} (1996) 3693
            [arXiv:hep-th/9603133];
          G.~Shiu and S.-H.H.~Tye, Phys.\ Rev.\ D {\bf 58} (1998) 106007
             [arXiv:hep-th/9805157];
          Z.~Kakushadze and S.-H.H.~Tye,
              Nucl.\ Phys.\ B {\bf 548} (1999) 180
             [arXiv:hep-th/9809147].


\bibitem{shape1}   K.~R.~Dienes,
         Phys.\ Rev.\ Lett.\  {\bf 88} (2002) 011601
         [arXiv:hep-ph/0108115].

\bibitem{shape2}   K.~R.~Dienes and A.~Mafi,
      Phys.\ Rev.\ Lett.\  {\bf 88} (2002) 111602
      [arXiv:hep-th/0111264].

\bibitem{shape3}   K.~R.~Dienes and A.~Mafi,
      Phys.\ Rev.\ Lett.\  {\bf 89} (2002) 171602
      [arXiv:hep-ph/0207009].


\bibitem{deconst}
        K.~Shiraishi, K.~Sakamoto and N.~Kan,
        arXiv:hep-ph/0209126.
       
\bibitem{toappear}  K.~R.~Dienes and A.~Mafi, to appear.
        

\end{thebibliography}
\section*{References}

\rightline{\it Nothing but papers, my 
    lord.\/\footnote{
    W.~Shakespeare, {\it  King Henry IV, Part I}\/:  II, iv.}}

\end{document}